\documentclass[twocolumn,aps,prd,superscriptaddress,nofootinbib,showpacs]{revtex4}%
\usepackage{graphicx}
\graphicspath{{fig/}{dia/}} \DeclareGraphicsExtensions{.eps}
\usepackage{subfig}
\usepackage{amsmath}
\usepackage{bm}
\usepackage[hypertex]{hyperref}

\usepackage{relsize}
\RequirePackage{xspace}

\newcommand{\jpsi}{J/\psi}
\newcommand{\chicj}{\chi_{cJ}}

\begin{document}

\title{QCD corrections to $\bm{e^+ e^- \to J/\psi(\psi(2S))+\chi_{cJ}}$ (J=0,1,2) at B factories}

\author{Kai Wang}
\affiliation{Department of Physics and State Key Laboratory of
Nuclear Physics and Technology, Peking University, Beijing 100871,
China}
\author{Yan-Qing Ma}
\affiliation{Department of Physics and State Key Laboratory of
Nuclear Physics and Technology, Peking University, Beijing 100871,
China}
\author{Kuang-Ta Chao}
\affiliation{Department of Physics
and State Key Laboratory of Nuclear Physics and Technology, Peking
University, Beijing 100871, China}\affiliation{Center for High
Energy Physics, Peking University, Beijing 100871, China}

\date{\today}

\begin{abstract}
We analytically calculate the cross sections of double charmonium
production in $e^+ e^- \to J/\psi(\psi(2S))\chi_{cJ}$ (J=0,1,2) at
next-to-leading order (NLO) in $\alpha_s$ in nonrelativistic QCD,
and confirm factorization of these processes. In contrast to
$\chi_{c0}$ production, for which the NLO correction is large and
positive, the NLO corrections for $\chi_{c1,2}$ production can be
negative,
resulting in decreased $K$ factors of 0.91 and 0.78 for J=1 and 2
respectively when $\mu = 2 m_{c}$. Consequently, the NLO QCD
corrections markedly enlarge the difference between cross sections
of $\chi_{c0}$ and $\chi_{c1,2}$.  This may explain why $e^+ e^- \to
J/\psi(\psi(2S))\chi_{c0}$ but not $e^+ e^- \to
J/\psi(\psi(2S))\chi_{c1,2}$ is observed experimentally. Moreover,
for $J/\psi(\psi(2S))\chi_{c1,2}$, the NLO QCD corrections
substantially reduce the $\mu$ dependence and lead to predictions
with small theoretical uncertainties.
\end{abstract}


\maketitle

\section{Introduction\label{sec:introduction}}

The study of heavy quarkonium inclusive production is important to
understanding the hadronization of heavy quarkonium. In the spirit of
nonrelativistic QCD (NRQCD) factorization \cite{Bodwin:1994jh}, it
is believed that both color-singlet and color-octet channels can
contribute to heavy quarkonium inclusive production.  However, due
to the lack of knowledge of color-octet long-distance matrix elements,
the inclusive production mechanism for heavy quarkonium is still not
fully understood even though complete next-to-leading order (NLO)
QCD corrections for most of these processes are available
\cite{Zhang:2006ay,Ma:2008gq,NLOpp,Ma:2010vd,Butenschoen:2010rq,NLOep}.
On the other hand, in heavy quarkonium exclusive production all
final state particles are targeted, and in some processes the
color-octet does not contribute at all, such as in double charmonium
production in $e^+ e^-$ annihilation. As a result, the study of
double charmonium exclusive production, among other kinds of production, may provide a
good opportunity to learn more about color-singlet mechanisms for
production and hadronization, without theoretical uncertainties from
color-octet contributions.

In 2002, the Belle Collaboration reported a surprisingly large double charmonium
production cross section\cite{Abe:2002rb}, which was later confirmed
by larger data samples\cite{Uglov:2004xa,Abe:2004ww} and also by
BaBar\cite{Aubert:2005tj}. The observed double charmonium production
cross sections for $e^+e^-\to J/\psi\eta_c(\chi_{c0})$ were much
larger than the leading order (LO) calculations\cite{Braaten:2002fi}
in NRQCD. Aside from attempted explanations by other models and
methods\cite{Braguta:2006py}, in the framework of NRQCD these
discrepancies were found to be essentially resolved by large NLO QCD
corrections\cite{Zhang:2005cha,Zhang:2008gp} and relativistic
corrections\cite{Bodwin:2006dn}.
It is puzzling that, although the cross section of $J/\psi
\chi_{c0}$ is observed to be very large, $\jpsi\chi_{c1,2}$ is not
seen. After all, at LO the cross section of $\jpsi\chi_{c0}$ is
larger than the total cross sections of $\jpsi\chi_{c1,2}$ by only a
factor of a little more than 2. Even with the most recent
data\cite{:2009nj}, signals of $\jpsi\chi_{c1,2}$ are still not
clearly seen. Recall that there was a similar situation for double
$\jpsi$ production. At LO, NRQCD predicts the $\jpsi\jpsi$ cross section
to be larger than that of $J/\psi\eta_c$ by a factor of
1.8\cite{Bodwin:2002fk}, but experimentally no evidence for double
$\jpsi$ production was observed. This puzzle was explained later by
finding a negative NLO QCD correction for double
$\jpsi$\cite{Gong:2008ce} and a positive NLO QCD correction for
$J/\psi\eta_c$\cite{Zhang:2005cha,Zhang:2008gp}. All these show that
the NLO QCD corrections can be very important for double charmonium
production. Therefore, it is necessary to examine the NLO QCD
corrections for $\jpsi\chi_{c1,2}$ as well as $\jpsi\chi_{c0}$
production in $e^+ e^-$ annihilation and to see whether the NLO QCD
corrections play important roles in understanding the nonobservation
of $\jpsi\chi_{c1,2}$ production at $B$ factories.

\begin{figure}
  \includegraphics[]{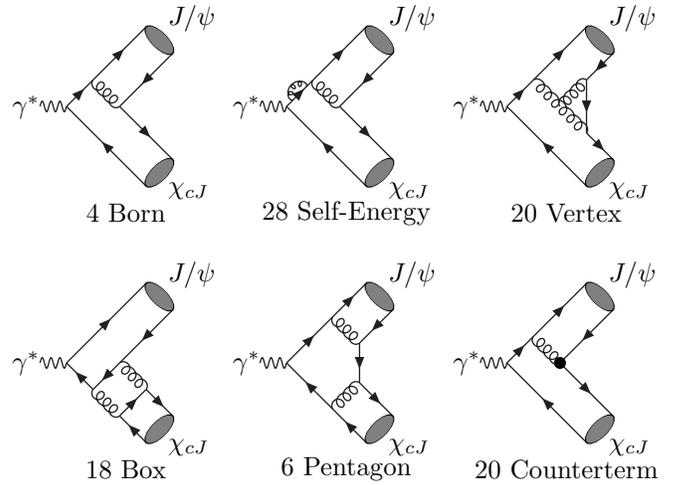}
  \caption{\label{fig:diagrams}
  Some representative Feynman diagrams
    for $e^+e^- \to J/\psi \chicj$.}
\end{figure}

The NLO QCD correction for $\jpsi\chi_{c0}$ has been calculated in
\cite{Zhang:2008gp}, and the difference between $\chi_{c0}$ and
$\chi_{c1,2}$ in the calculation is the summation over
polarizations. However, the NLO QCD correction for
$\jpsi\chi_{c1,2}$ is indeed a nontrivial work even though that for
$\jpsi\chi_{c0}$ has been achieved. This is because $\chi_{c0}$ is
a scalar particle, so the summation over its spin and angular
polarizations can be done at the amplitude level, which makes its
Lorentz structure much simpler than that of $\chi_{c1,2}$. However, the
method used to calculate $\jpsi\chi_{c0}$ in \cite{Zhang:2008gp} can
hardly be adopted here to calculate $\jpsi\chi_{c1,2}$. Furthermore,
the method in \cite{Zhang:2008gp} gives only numerical results, while
an analytical expression is really important for analyzing the details
of the result.  Fortunately, using our recently developed method,
which has been used to do the complete NLO corrections for heavy
quarkonium hadroproduction \cite{Ma:2010vd}, analytical expressions
of the production cross sections of $e^+ e^- \to
J/\psi(\psi(2S))\chi_{cJ}$ at NLO can be conveniently achieved.

\section{{Calculation}\label{sec:{Calculation}}}

In the following, we briefly describe our calculation. We use {\tt
FeynArts}\cite{feynarts} to generate Feynman diagrams and Feynman
amplitudes.  Some representative Feynman diagrams for this process
are shown in Fig.~\ref{fig:diagrams}. There are generally
ultraviolet(UV), infrared(IR), and Coulomb singularities.
Conventional dimensional regularization (CDR) with $D=4-2\epsilon$
is adopted to regularize them.  The UV-divergences from self-energy
and triangle diagrams are removed by renormalization. The Coulomb
singular terms are factored into the $J/\psi$ and $\chicj$ wave
functions. Differing from S-wave charmonium production, where IR
singularities can be canceled to ensure  factorization, there are
generally nonfactorizable IR divergences in P-wave charmonium
exclusive production (such as $B \to \chicj K$\cite{Song:2002mh})
due to the nonvanishing relative momentum between the heavy quark
and the antiquark. As previously discussed in \cite{Zhang:2008gp}, it is
the existence of the associated S-wave state $\jpsi$ that avoids
topologically nonfactorizable soft interactions and leads to
cancellation of IR singularities. This argument is generalized in
\cite{Bodwin:2008nf} to prove the factorization theorem of heavy
quarkonium exclusive production to all orders in $\alpha_s$. In the
present work, we find that the IR singularities are indeed canceled
between different diagrams, which further confirms factorization of
$e^+ e^- \to J/\psi\chi_{cJ} (J=0,1,2)$.

To perform the calculation, two different methods are used,
resulting in two completely independent computer codes. The results
are found to agree with each other with high precision. Furthermore,
our result for $\chi_{c0}$ is basically consistent with that in
\cite{Zhang:2008gp} \footnote{There is an error in the numerical
calculation in \cite{Zhang:2008gp}. Our result is consistent with it
only after correcting the error in \cite{Zhang:2008gp}.}.

One of our calculations is based on the {\tt Mathematica} package {\tt
FeynCalc}\cite{Mertig:1990an}. We separate the soft singularities in
the virtual corrections using the method used in \cite{Dittmaier:2003bc},
and then we treat the singular part analytically and the finite part
numerically. To calculate the finite part, we use the traditional
method \cite{Passarino:1978jh} to reduce tensor loop integrations to
scalar functions, which are then calculated numerically by {\tt
LoopTools}\cite{looptools}. When derivatives of finite scalar
functions are needed, we perform them numerically. This method
involves large numerical cancellations, so we use quadruple
precision in the calculation to guarantee the result to be reliable.
The other one of our calculations is based on our self-written {\tt Mathematica} codes.
The derivatives are performed before tensor reductions and loop
integrations, so it avoids the derivatives of scalar functions.
Then we decompose tensor integrals and reduce scalar functions to a
fundamental set using integrate-by-part-based recursion relations.
The fundamental set consists of only one, two, and three point
scalar functions in this work.  We use the analytical expressions in
\cite{Ellis:2007qk} for divergent scalar functions, while finite
scalar functions are calculated  analytically with the methods in
\cite{'tHooft:1978xw}. We use {\tt QCDLoop}\cite{Ellis:2007qk} to
check our analytical expressions of scalar functions. Final
analytical results for the cross sections are given in the Appendix.

\begin{figure*}[hbtp]
\centering \subfloat[$e^{+}+e^{-}\to\jpsi+\chi_{c0}$]{
  \begin{minipage}[t]{.33\linewidth}
    \includegraphics[width=\linewidth]{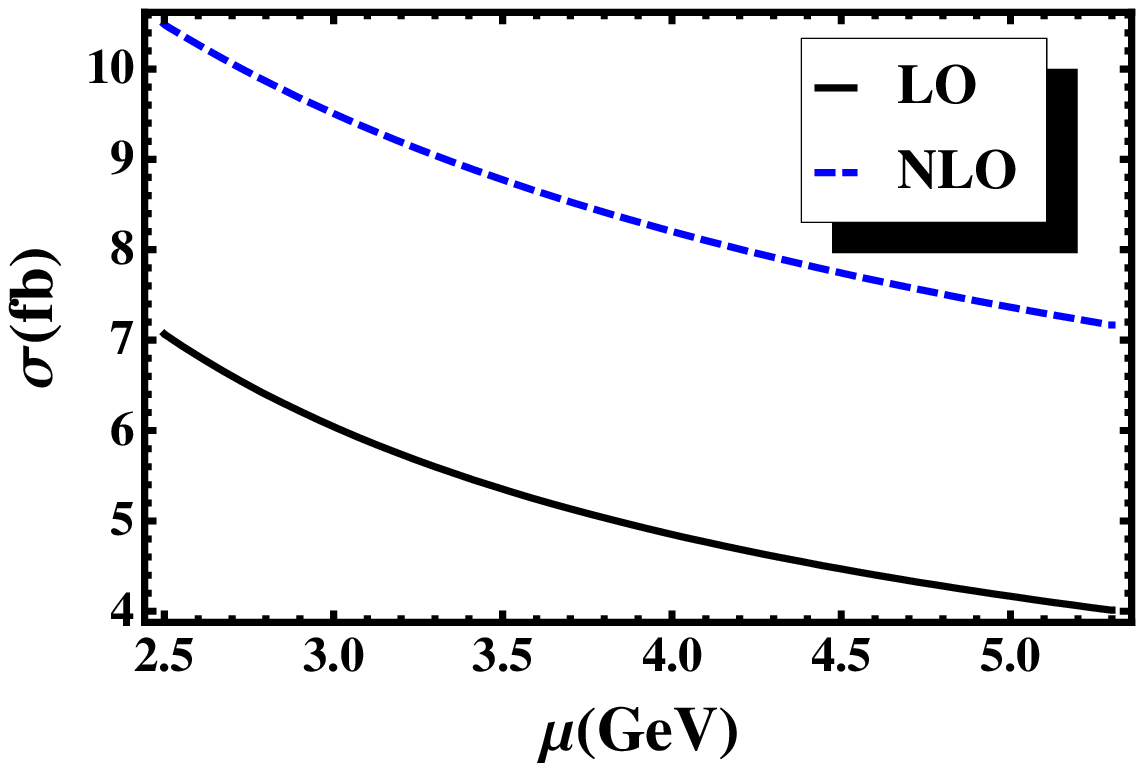}
  \end{minipage}
} \subfloat[$e^{+}+e^{-}\to\jpsi+\chi_{c1}$]{
  \begin{minipage}[t]{.33\linewidth}
    \includegraphics[width=\linewidth]{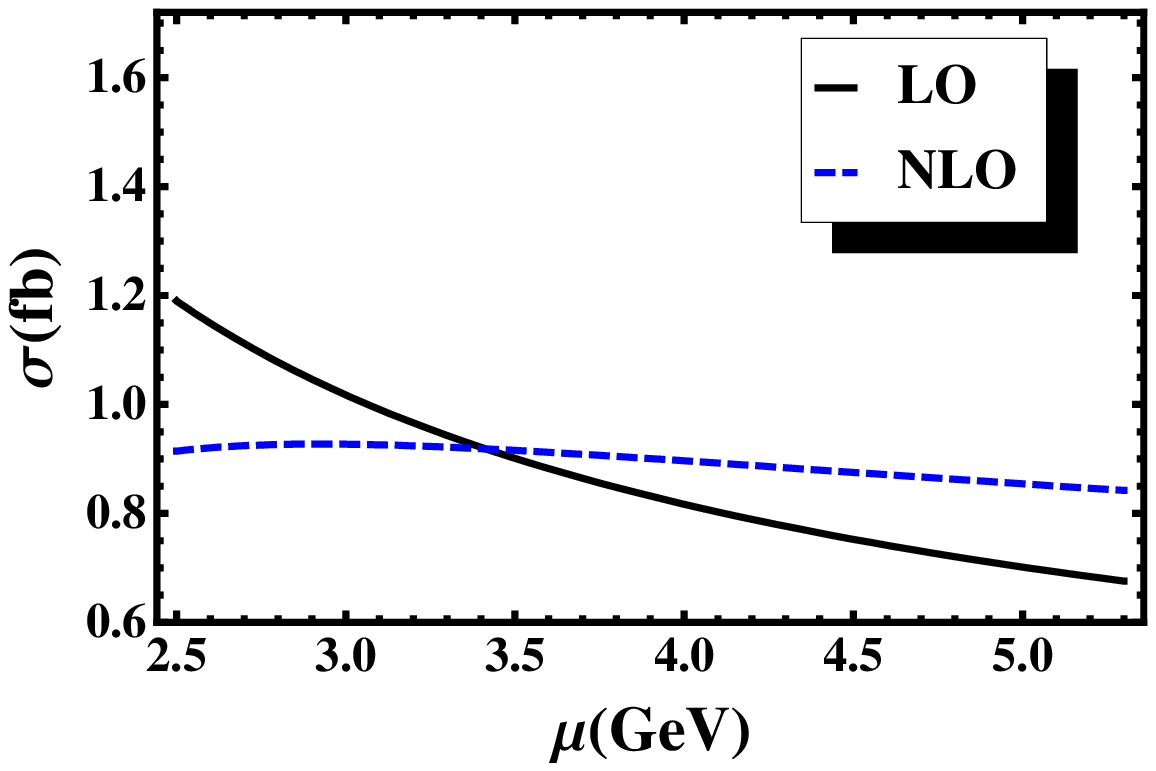}
  \end{minipage}
} \subfloat[$e^{+}+e^{-}\to\jpsi+\chi_{c2}$]{
  \begin{minipage}[t]{.33\linewidth}
    \includegraphics[width=\linewidth]{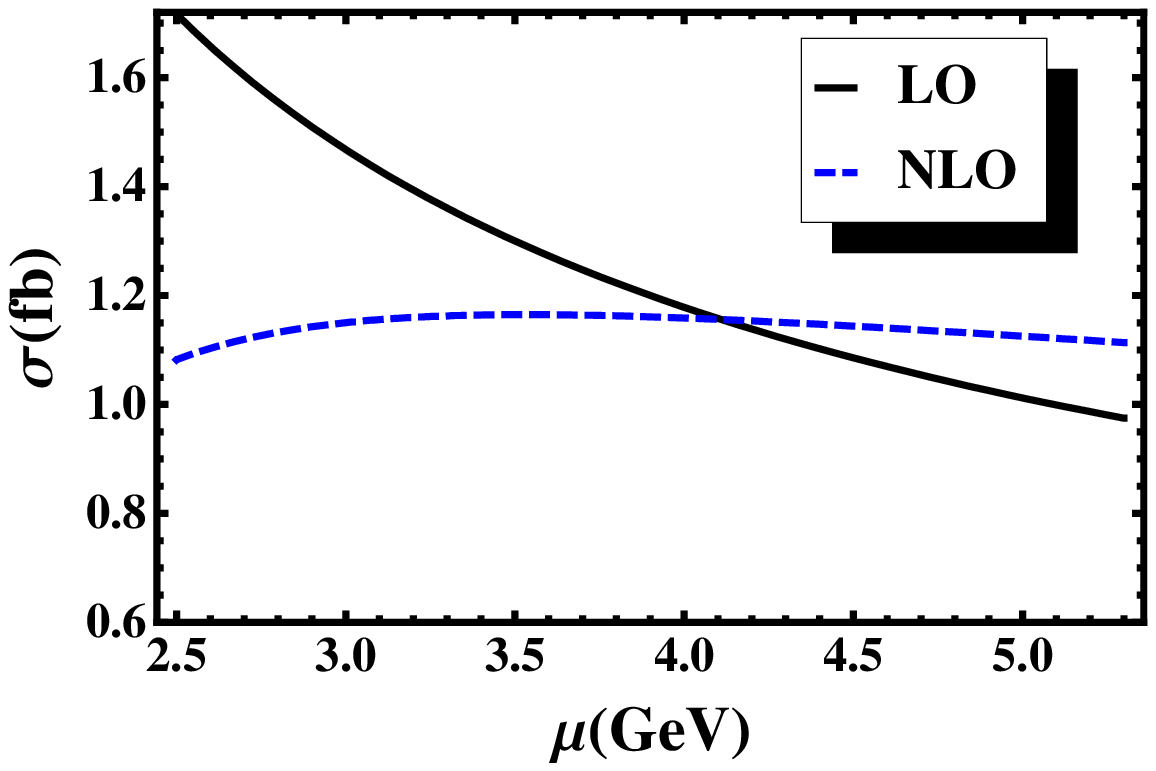}
  \end{minipage}
} \caption{\label{fig:mudep} Cross sections of $e^+ e^- \to J/\psi
  \chicj$ as functions of the renormalization scale $\mu$ at LO and
  NLO in $\alpha_s$ with $\sqrt{s}=10.6{\rm GeV}$, $\Lambda=338{\rm
    MeV}$, and $m_{c}=1.5{\rm GeV}$.}
\end{figure*}

\begin{figure*}[hbtp]
\centering \subfloat[$e^{+}+e^{-}\to\jpsi+\chi_{c0}$]{
  \begin{minipage}[t]{.33\linewidth}
    \includegraphics[width=\linewidth]{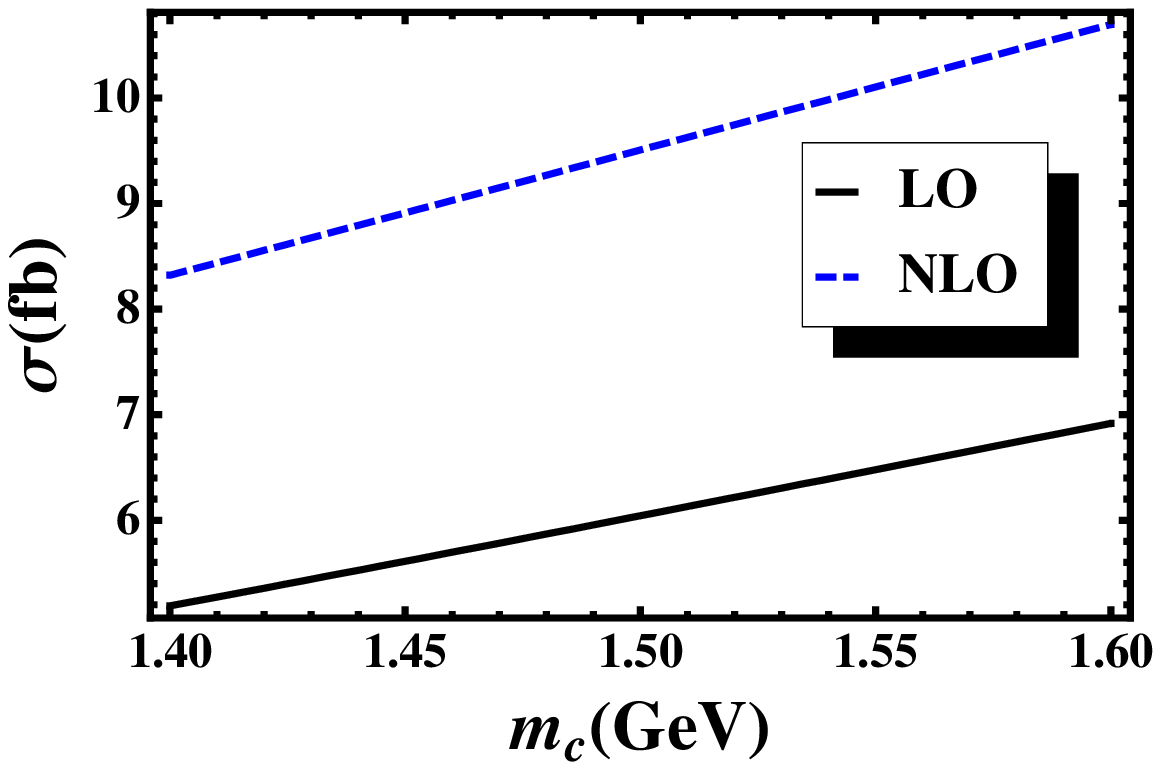}
  \end{minipage}
} \subfloat[$e^{+}+e^{-}\to\jpsi+\chi_{c1}$]{
  \begin{minipage}[t]{.33\linewidth}
    \includegraphics[width=\linewidth]{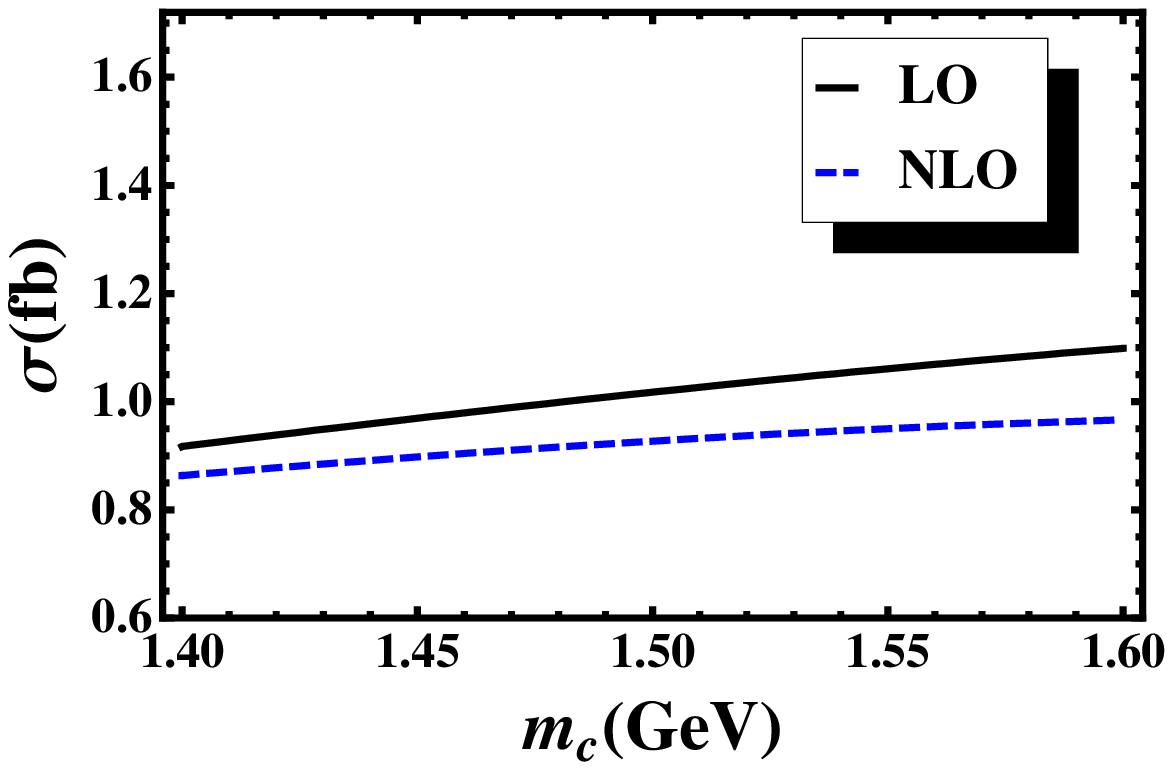}
  \end{minipage}
} \subfloat[$e^{+}+e^{-}\to\jpsi+\chi_{c2}$]{
  \begin{minipage}[t]{.33\linewidth}
    \includegraphics[width=\linewidth]{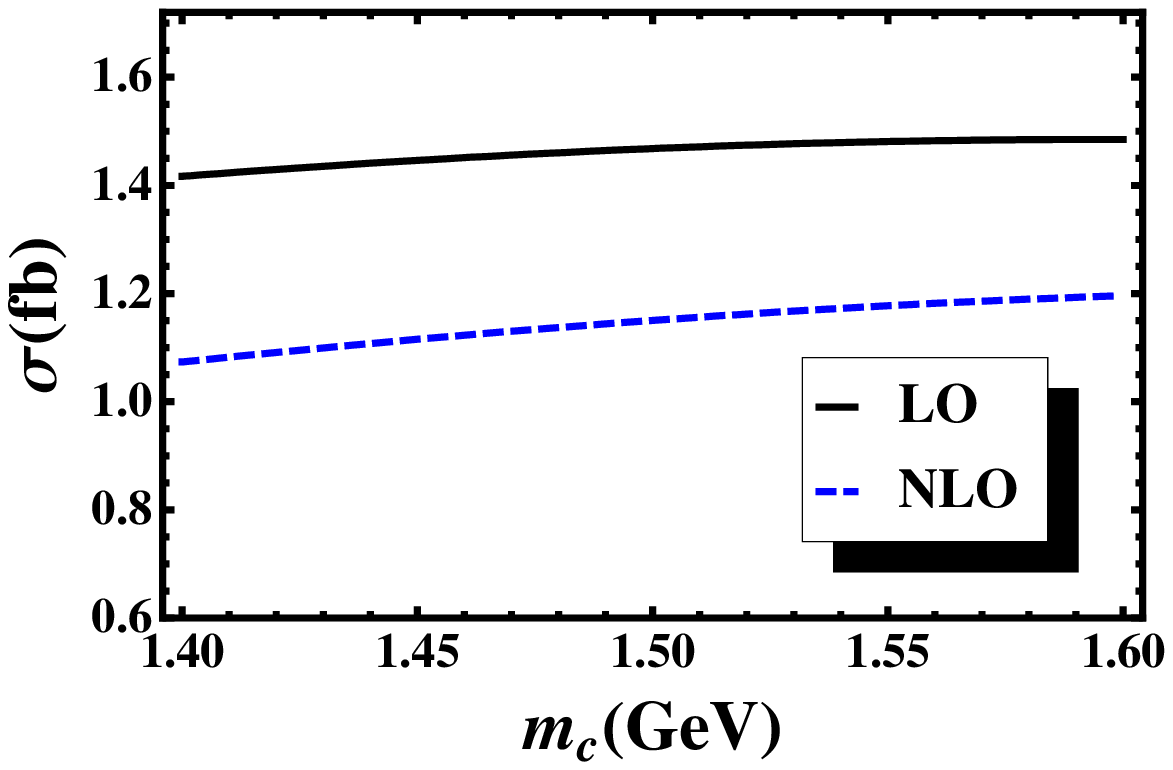}
  \end{minipage}
} \caption{\label{fig:mcdep} Cross sections of $e^+ e^- \to J/\psi
  \chicj$ as functions of the charm quark mass $m_{c}$ at LO and NLO in
  $\alpha_s$ with $\sqrt{s}=10.6{\rm GeV}$, $\Lambda=338{\rm MeV}$,
  and $\mu = 2 m_{c}$.}
\end{figure*}


\section{Result and discussion}

In the numerical calculation, we set $\sqrt{s}=10.6$ GeV,
$\Lambda_{\overline{MS}}^{(4)}=338$ MeV, and $m_{c}=1.5$ GeV. The
values of wave functions squared at the origin (or their
derivatives) are extracted from the leptonic width of $\jpsi$ and
$\psi(2S)$ and the two-photon width of
$\chi_{c2}$\cite{Amsler:2008zz}
at NLO in $\alpha_s$: 
\begin{align}
  \label{eq:6}
  &\Gamma_{\psi(nS)\to e^+e^-}=\frac{4\alpha^{2}}{9 m_{c}^{2}}
  \left(1-\frac{16}{3}\frac{\alpha_{s}}{\pi}\right)\lvert R_{nS}(0)\rvert^{2},\\
  \label{eq:rp2chic2}
  &\Gamma_{\chi_{c2}(nP)\to \gamma\gamma}=\frac{64\alpha^{2}}{45 m_{c}^{4}}
  \left(1-\frac{16}{3}\frac{\alpha_{s}}{\pi}\right)\lvert
  R'_{nP}(0)\rvert^{2}.
\end{align}
Setting $\alpha_{s}=0.259$ and $\alpha=\frac{1}{137}$ we get
\begin{align}
  \label{eq:7}
  \lvert R_{1S}(0)\rvert^{2}&=0.418{\rm GeV}m_{c}^{2},\\
  \lvert R_{2S}(0)\rvert^{2}&=0.179{\rm GeV}m_{c}^{2},\\
  \lvert R'_{1P}(0)\rvert^{2}&=0.0116{\rm GeV}m_{c}^{4}.
\end{align}
Note that the values of wave functions squared at the origin here
depend on charm quark mass $m_{c}$. This treatment can largely
cancel the $m_{c}$ dependence in the short-distance coefficients and
significantly reduce the theoretical uncertainty due to the choice
of $m_{c}$.

In Table \ref{tab:kfac}, we give the ratio of cross section at NLO
to that at LO (the K factor). Different from the $\chi_{c0}$
production, where the K factor is much larger than 1, the K factors
for $\chi_{c1,2}$ are small, and the NLO corrections are small and
even negative when $\mu=2 m_{c}$. Therefore, the gap of cross
sections between $\chi_{c0}$ and $\chi_{c1,2}$ is further enlarged
at NLO, which gives an explanation for why only
$J/\psi(\psi(2S))\chi_{c0}$ production is observed. In
Fig.~\ref{fig:mudep}, we show the cross sections at LO and NLO as
functions of the renormalization scale $\mu$. We find that, although
$\mu$ dependence for $\chi_{c0}$ is large at both LO and NLO, QCD
corrections substantially reduce the $\mu$ dependence for
$\chi_{c1,2}$. The dependence of charm quark mass is shown in
Fig.~\ref{fig:mcdep}, where we find the dependence for $\chi_{c1,2}$
is also weaker than that for $\chi_{c0}$. Based on
Figs.~\ref{fig:mudep} and \ref{fig:mcdep}, we conclude that the
NLO predictions for $\chi_{c1,2}$ production have small theoretical
uncertainties; therefore, they can be used to precisely test the
production mechanism when they can be measured in future experiment.

\begin{table}[htbp]
  \centering
  \begin{tabular}{lcccc}
    \toprule
    &$\alpha_s(\mu)$&$J/\psi +\chi_{c0}$&$J/\psi +\chi_{c1}$&$J/\psi +\chi_{c2}$\\
    \hline
    $\mu = 2 m_{c}$&0.259&1.57&0.91&0.78\\
    $\mu = \sqrt{s}/2$&0.211&1.79&1.25&1.14\\
    \botrule
  \end{tabular}
  \caption{The K factor of our QCD corrections, with $\sqrt{s}=10.6{\rm GeV}$, $\Lambda=338{\rm
      MeV}$, and $m_{c}=1.5{\rm GeV}$.}
  \label{tab:kfac}
\end{table}


The comparison of experimental data with our theoretical predictions
is shown in Table \ref{tab:exth}, where we vary $m_{c}=1.5 \pm 0.1$
GeV to estimate the theoretical uncertainties. Our results are
consistent with all data except $e^{+}e^{-}\to\psi(2S)\chi_{c0}$.
The Belle Collaboration's observation that the cross section of
$e^{+}e^{-}\to\psi(2S)\chi_{c0}$ is bigger than that of
$e^{+}e^{-}\to\jpsi\chi_{c0}$ cannot be explained in NRQCD at LO in
$v$, the relative velocity of charm quark and anticharm quark,
because the only difference between $\psi(2S)$ and $\jpsi$ at LO in
$v$ is the wave functions squared at the origin, which are well
estimated.  Perhaps the relativistic corrections might give some
hint, but the large errors in experiment should be reduced before
any definite conclusion can be drawn.

\begin{table*}[htbp]
  \centering
  \begin{tabular}{ccccccc}
    \toprule
    & Belle
    & BaBar
    & LO result
    & NLO result
    & LO result
    & NLO result\\
    & $\sigma \times \mathcal{B}_{>2(0)}$\cite{Abe:2004ww}
    & $\sigma \times \mathcal{B}_{>2}$\cite{Aubert:2005tj}
    & ($\mu=2m_{c}$)
    & ($\mu=2m_{c}$)
    & ($\mu=\sqrt{s}/2$)
    & ($\mu=\sqrt{s}/2$)\\
    \hline
    $\sigma(J/\psi +\chi_{c0})$ & $6.4 \pm 1.7 \pm 1.0 $&$10.3 \pm 2.5 ^{+1.4}_{-1.8} $&
    $6.0 \pm 0.9$ &$9.5 \pm 1.2$&$4.0 \pm 0.8$&$7.2 \pm 1.2$\\
    $\sigma(J/\psi +\chi_{c1})$ &-&-&
    $1.02^{+0.08}_{-0.10}$ &$0.93^{+0.04}_{-0.07}$&$0.68^{+0.09}_{-0.10}$&$0.84^{+0.07}_{-0.09}$\\
    $\sigma(J/\psi +\chi_{c2})$ &-&-&
    $1.47^{+0.01}_{-0.05}$ &$1.15^{+0.05}_{-0.08}$&$0.97^{+0.07}_{-0.08}$&$1.11^{+0.05}_{-0.07}$\\
    $\sigma(J/\psi +\chi_{c1})+\sigma(J/\psi +\chi_{c2})$
    & $<$5.3 at 90\% C.L.&-&
    $2.49^{+0.09}_{-0.16}$&$2.08^{+0.08}_{-0.14}$&$1.65^{+0.16}_{-0.19}$&$1.96^{+0.11}_{-0.17}$\\
    $\sigma(\psi(2S) +\chi_{c0})$& $12.5 \pm 3.8 \pm 3.1 $&-&
    $2.6 \pm 0.4$&$4.1 \pm 0.5$&$1.7^{+0.4}_{-0.3}$&$3.1 \pm 0.5$\\
    $\sigma(\psi(2S) +\chi_{c1})$&-&-&
    $0.44^{+0.03}_{-0.05}$&$0.40^{+0.01}_{-0.03}$&$0.29 \pm 0.04$&$0.36^{+0.03}_{-0.04}$\\
    $\sigma(\psi(2S) +\chi_{c2})$&-&-&
    $0.63^{+0.01}_{-0.02}$&$0.49^{+0.02}_{-0.03}$&$0.42^{+0.02}_{-0.04}$&$0.48^{+0.02}_{-0.04}$\\
    $\sigma(\psi(2S) +\chi_{c1})+\sigma(\psi(2S) +\chi_{c2})$
    & $<$8.6 at 90\% C.L.&-&
    $1.06^{+0.05}_{-0.06}$&$0.89^{+0.04}_{-0.06}$&$0.71^{+0.06}_{-0.08}$&$0.84^{+0.05}_{-0.07}$\\
    \botrule
  \end{tabular}
  \caption{Comparison of our predicted cross sections with experiments
    at B factories in units of fb.
    The errors of our theoretical predictions are only from variations of
    the charm quark mass $m_{c}=1.5 \pm 0.1 {\rm GeV}$.
    The experimental data are cross sections
    times branching fractions for
    $\chi_{cJ} $ decay into more than 2 charged tracks, while the Belle
    data of $\psi(2S) +\chi_{cJ}$ in \cite{Abe:2004ww} correspond to
    $\chi_{cJ} $ decay into at least 1 charged track. }
  \label{tab:exth}
\end{table*}


\section{Summary}

In NRQCD, using an improved method, we get compact analytical
expressions of  double charmonium production cross sections of $e^+
e^- \to J/\psi(\psi(2S))\chi_{cJ}$ (J=0,1,2) at NLO in $\alpha_s$
and LO in $v$. Moreover, we further confirm  factorization of these
processes. With $\sqrt{s}=10.6{\rm GeV}$, $m_{c}=1.5{\rm GeV}$, and
$\mu = 2 m_{c}$, we find that the cross section for $\chi_{c0}$ is
enhanced by a $K$ factor of 1.57, while the cross sections of
$\chi_{c1,2}$ are decreased with $K$ factors of 0.91 and 0.78
respectively. The large positive NLO correction to $\chi_{c0}$ and
negative NLO corrections to $\chi_{c1,2}$ markedly enlarge the
difference between the cross sections of $\chi_{c0}$ and
$\chi_{c1,2}$ and provide an explanation for  the phenomenon that
only $J/\psi(\psi(2S))\chi_{c0}$ but not
$J/\psi(\psi(2S))\chi_{c1,2}$ production is observed. Considering
the substantially reduced  $\mu$ dependence, and also the small
$m_{c}$ dependence, the NLO predictions for $e^+ e^- \to
J/\psi(\psi(2S))\chi_{c1,2}$ are more precise than for many other
processes and may be used to test the heavy quarkonium production
mechanism in the future. However, the predicted cross section for
$\psi(2S)\chi_{c0}$ production is much smaller than the data. In
order to clarify this problem, we suggest that both more careful
measurement be performed for this process and further theoretical
investigations be made in the future.

\begin{acknowledgments}
We thank Y.J. Zhang for helpful discussions. This work was supported
by the National Natural Science Foundation of China (Contract No.11021092 and
No.11075002) and the Ministry of Science and Technology of China
(Contract No.2009CB825200).
\end{acknowledgments}

{\it{Note added in proof.}}--After this work was submitted for publication, a
preprint \cite{Dong:2011fb} appeared in which the obtained result is consistent with ours.

\appendix*
\section{Analytical Result}
Here we present the analytical result of our calculation.  For
brevity of expression, we define
 \begin{equation}
 \begin{split}
   &a=\tfrac{\sqrt{s}}{4 m_c},b=\sqrt{a^2-1},c=\sqrt{4 a^2-1},d=8
   a^2+1,\\& e=2 a^2-1,f=8 a^4+4 \left(2 a^2+1\right) a b+1,\\& g=2 a
   b+1,h=8 a^4-8 a^2-4 a b e+1.
 \end{split}
 \end{equation}
 The LO cross sections for $e^{+}+e^{-}\to\jpsi+\chicj$ are
 \begin{equation}
   \sigma^{{\rm LO}}_{\chicj}=
   \frac{b \pi \alpha ^2 \alpha_s^2 } {3888 a^{15} m_c^{10}} \lvert
   R_{S}(0)\rvert^{2} \lvert R'_{P}(0)\rvert^{2}M_{J}^{{\rm LO}},
 \end{equation}
 where $M_{J}^{{\rm LO}}$ for $\chi_{c0}$, $\chi_{c1}$, and $\chi_{c2}$ are given by
 \begin{align}
   M_{0}^{{\rm LO}}&=16 a^8+728 a^6-428 a^4+38 a^2+9,\\
   M_{1}^{{\rm LO}}&=192 a^6-288   a^4+78 a^2+27,\\
   M_{2}^{{\rm LO}}&=32 a^8+160 a^6-376 a^4+154 a^2+45.
 \end{align}
 The NLO cross sections are
 \begin{equation}
   \label{eq:2}
   \sigma^{{\rm NLO}}_{\chicj}=\sigma^{{\rm LO}}_{\chicj}
   \left(1 +\frac{\alpha_s\text{Re} M_{J}^{{\rm NLO}}}{72 a^3 b d^3 e\pi
     M_{J}^{{\rm LO}} } \right),
 \end{equation}
 where $M_{J}^{{\rm NLO}}$ (only real parts are guaranteed to be correct)
 are given by
 \begin{widetext}

 \begin{equation}
   \label{eq:3}
   \begin{split}
     M_{0}^{{\rm NLO}}=& -24 \big(256 a^{12}+7840 a^{10}-14284 a^8+8340
     a^6-1943 a^4+69 a^2+30\big) a^2 d^3 l_1+96 \big(32 a^8-1420
     a^6\\&+1710 a^4-321 a^2-45\big) a^4 d^3 e l_2 -24 \big(64
     a^{10}+2904 a^8+636 a^6-784 a^4+42 a^2+9\big) a^2 d^3 e l_3
     \\&-216 \big(40 a^4-4 a^2-3\big) a^6 d^3 l_4+12 \big(256
     a^{12}+19328 a^{10}-25596 a^8+10840 a^6-1537 a^4-45
     a^2\\&+21\big)a^2 d^3 l_5 +12 \big(19328 a^{10}-33644
     a^8+17748 a^6-2867 a^4-111 a^2+30\big) a^2 d^3 l_6+24
     \big(96 a^{10}\\&+768 a^8-812 a^6+430 a^4-66 a^2-9\big) a^2 d^3 e
     l_7+4 \big(13824 a^{12}+273152 a^{10}-152936 a^8+20260
     a^6\\&+4376 a^4-1059 a^2-108\big)a b d^2 e+2 \big(192 a^{10}+11200
     a^8-13254 a^6+3883 a^4-81 a^2-48\big)a^2 d^3 e \pi ^2\\& -8
     \big(6002688 a^{12}+1326848 a^{10}-2958336 a^8-147576 a^6+198772
     a^4+31524 a^2+1377\big)a^3 b d e \ln 2\\& -8 \big(3035136
     a^{12}+3154432 a^{10}-2263296 a^8-276912 a^6+133688 a^4+24456
     a^2+1125\big)a^3 b d e \ln a\\& -16 \big(8 a^{12}+656
     a^{10}-2040 a^8+4108 a^6-1796 a^4+126 a^2+27\big) d^3 e \ln
     (a+b)+16 \big(11752 a^8\\&+5628 a^6-8364 a^4+1430 a^2+147\big)a^2
     b c d^3 e \ln (2 a+c) +600 \big(16 a^8+728 a^6-428 a^4+38
     a^2\\&+9\big) a^3 b d^3 e \ln(\mu/m_{c} ),
   \end{split}
 \end{equation}
 \begin{equation}
   \label{eq:4}
   \begin{split}
     M_{1}^{{\rm NLO}}= & +36 \big(896 a^8+160 a^6-1200 a^4+125 a^2+60\big)
     a^2 d^3 e l_1 + 144 \big(848 a^6+358 a^4-471 a^2-81\big) a^4
     d^3 e l_2\\& -72 \big(1648 a^8 -550 a^6-436 a^4+18
     a^2+9\big) a^2 d^3 e l_3 +18 \big(704 a^8-1976 a^6+1270
     a^4+73 a^2\\&-42\big)a^2 d^3 e l_5 -18 \big(4288 a^8+1592
     a^6-3084 a^4-199 a^2+60\big) a^2 d^3 e l_6 -72 \big(160
     a^8-100 a^6-162 a^4\\&+42 a^2+9\big) a^2 d^3 e l_7 +12
     \big(47104 a^{12}-164608 a^{10}+83232 a^8+36524 a^6-2732 a^4-1599
     a^2\\&-108\big) a b d e -3 \big(4928 a^8+1192 a^6-3732 a^4-31
     a^2+96\big) a^2 d^3 e \pi ^2 -72 \big(524288 a^{14}-2965504
     a^{12}\\&-1094656 a^{10}+1111040 a^8+689584 a^6+145600 a^4+13334
     a^2+459\big) a^3 b e \ln 2 -72 \big(262144 a^{14}\\&-311296
     a^{12}-1196032 a^{10}+335616 a^8+419712 a^6+104776 a^4+10404
     a^2+375\big) a^3 b e \ln a\\& -48 \big(112 a^{10}-780 a^8+1173
     a^6-544 a^4+45 a^2+27\big) d^3 e \ln (a+b) +48 \big(384 a^8-592
     a^6-946 a^4\\&+1010 a^2+147\big) a^2 b c d^3 e \ln (2 a+c) +1800
     \big(64 a^6-96 a^4+26 a^2+9\big) a^3 b d^3 e \ln (\mu/m_{c} ),
   \end{split}
 \end{equation}
 \begin{equation}
   \label{eq:5}
   \begin{split}
     M_{2}^{{\rm NLO}}= & -12 \big(1024 a^{12}+4864 a^{10}+896 a^8+11592
     a^6-7070 a^4-249 a^2+300\big) a^2 d^3 l_1 +48 \big(128
     a^8-1528 a^6\\&-4578 a^4-1833 a^2-369\big) a^4 d^3 e l_2 -24
     \big(128 a^{10}+600 a^8-7794 a^6-1862 a^4+30 a^2+45\big) a^2 d^3 e
     l_3\\& -648  \big(4 a^2+1\big) a^4 b^{2} d^3 l_4
     +6 \big(1024 a^{12}+13376 a^{10}-16224 a^8+17908 a^6-5416
     a^4-867 a^2\\&+210\big) a^2 d^3 l_5 +6 \big(17600
     a^{10}+31408 a^8+7944 a^6-11450 a^4-1725 a^2+300\big) a^2 d^3
     l_6  +24 \big(192 a^{10}\\&+720 a^8-100 a^6+230 a^4-150
     a^2-45\big) a^2 d^3 e l_7 +4 \big(147456 a^{14}+825344
     a^{12}-1544320 a^{10}\\&+468720 a^8 +189436 a^6-20996 a^4-8643
     a^2-540\big) a b d e +\big(768 a^{10}+11680 a^8+19704 a^6+15160
     a^4\\&+633 a^2-480\big) a^2 d^3 e \pi ^2 -8 \big(11304960
     a^{14}+25518080 a^{12}+43710976 a^{10}+31362816 a^8+11667664
     a^6\\&+2211824 a^4+199902 a^2+6885\big) a^3 b e \ln 2 -8 \big(5898240
     a^{14} +17588224 a^{12}+12001280 a^{10}+13653888 a^8\\&+7211936
     a^6+1604320 a^4+156660 a^2+5625\big) a^3 b e \ln a +16 \big(1712
     a^{12}+1016 a^{10}-3984 a^8+469 a^6\\&+1264 a^4-387 a^2-135\big)
     d^3 e \ln (a+b) +16 \big(1808 a^8+3192 a^6+5286 a^4+4486
     a^2\\&+735\big) a^2 b c d^3 e \ln (2 a+c) +600 \big(32 a^8+160
     a^6-376 a^4+154 a^2+45\big) a^3 b d^3 e \ln (\mu/m_{c} ),
   \end{split}
 \end{equation}
 where $l_{1}$--$l_{7}$ are combinations of dilogarithms:
 \begin{equation}
   \label{eq:1}
   \begin{split}
     l_1&=\text{Li}_2\left(1+b/a\right),l_2=\text{Li}_2\left(1-b/a\right),
     l_3=\text{Li}_2(e+2 a   b)-\text{Li}_2(e-2 a   b),\\
     l_4&=2 \text{Li}_2(g)-\text{Li}_2\left(d g/f\right)
     +\text{Li}_2\left(1/h\right)-\text{Li}_2\left(g/h\right)-\text
     {Li}_2(h)
     +\text{Li}_2\left(d   h/f\right),\\
     l_5&=\text{Li}_2\left[\frac{(a+b)^2}{e+b c}\right]
     -\text{Li}_2\left[\frac{(a-b)^2}{e+b c}\right]
     +\text{Li}_2\left[\frac{(a+b)^2}{e-b c}\right]
     -\text{Li}_2\left[\frac {(a-b)^2}{e-b c}\right],\\
     l_6&=\text{Li}_2\left(4 a^2-3-\frac{b
         c^2}{a}\right)+\text{Li}_2\left(\frac{4 a^3+5 a-b c^2}{a
         d}\right)-\text{Li}_2\left(\frac{d}{32 a^4-24 a^2-8 b c^2
         a+1}\right),\\
     l_7&=\text{Li}_2\left(\frac{a^2+ a
         b+1}{2}\right)-\text{Li}_2\left(\frac{a^2- a b+1}{2}\right)
     +\text{Li}_2\left(\frac{a^2+ a b+1}{2 a^2}\right)
     -\text{Li}_2\left(\frac{a^2-  a b+1}{2 a^2}\right)\\
     &-2 \text{Li}_2\left(1-\frac{b}{2 a}\right)+\text{Li}_2\left(a^2-b
       a-\frac{b}{2 a}\right)+\text{Li}_2\left[\frac{14 a^3+4 a-\left(2
           a^2+1\right) b}{2 a
         d}\right]-\text{Li}_2\left(\frac{f}{d}\right).
   \end{split}
 \end{equation}
 \end{widetext}


\end{document}